# Dynamic modulation of thermal emission - a tutorial

Michela F. Picardi,[1] Kartika N. Nimje,[1] and Georgia T. Papadakis[1, a)]
*ICFO - Institut de Ciencies Fotoniques, The Barcelona Institute of Science and Technology,*
*Castelldefels (Barcelona) 08860, Spain*

Thermal emission is typically associated with a blackbody at a temperature above absolute zero, which exchanges energy with its environment in the form of radiation. Blackbody thermal emission is largely incoherent both spatially and temporally. Using principles in nanophotonics, thermal emission with characteristics that differ considerably from those of a blackbody have been demonstrated. In particular, by leveraging intrinsic properties of emerging materials or via nanostructuring at the wavelength or sub-wavelength scale, one can gain control over the directionality, temporal coherence, and other more exotic properties of thermal radiation. Typically, however, these are fixed at the time of fabrication. Gaining *dynamic* control of thermal emission requires exploiting external mechanisms that actively modulate radiative properties. Numerous applications can benefit from such thermal emission control, for example in solar energy harvesting, thermophotovoltaic energy conversion, radiative cooling, sensing, spectroscopy, imaging and thermal camouflage. In this tutorial, we introduce thermal emission in two domains: the far-field, and the near-field, and we outline experimental approaches for probing thermal radiation in both ranges. We discuss ways for tailoring the spatial and temporal coherence of thermal emission and present available mechanisms to actively tune these characteristics.

## I. INTRODUCTION: THERMAL RADIATION

From the sun itself, to the discovery of fire, all the way to incandescent light bulbs, radiation generated by hot objects has been the world's primary source of illumination consistently throughout history. While radiation from burning objects was already known to our cavemen ancestors, the fact that *all* bodies emit thermal radiation if their temperature is above 0 K was only realized in the nineteenth century. Only a small portion of this radiation is visible to the unaided human eye, and it originates from bodies at very high-temperatures, such as the sun. At a temperature of nearly 6000 K, the sun emits light mainly at visible and near-infrared (IR) frequencies. At lower, terrestrial temperatures, near 300 K, from Planck's law of thermal radiation[1], the peak of a blackbody's thermal emission spectrum shifts to the mid-IR range, corresponding to wavelengths in the range of 5-20 $\mu$m.

The first connection between light and its thermal properties is tied to the discovery of IR radiation by William Herschel in 1800[2]. Observing the spectrum of sunlight, dispersed through a prism, Herschel utilized a series of thermometers to measure the temperature of the radiation corresponding to each color. He then placed a thermometer *beyond* the visible red end and recorded a high-temperature, unveiling the existence of invisible radiation carrying thermal energy. Driven by the interest to explain astronomical observations, numerous attempts were subsequently made to describe thermal radiation, a notable one resulting in the Rayleigh-Jeans law[3], which predicts the spectral radiance of a blackbody to be proportional to $\lambda^{-4}$. While constituting a good approxima-tion for long wavelengths, the Rayleigh-Jeans law hints at an *ultraviolet catastrophe* at smaller wavelengths, for which the spectral radiance would diverge. Max Planck resolved the paradox by introducing his famous law of thermal radiation[1]. Indeed, Planck's theory reduces to Rayleigh-Jeans law for large values of $\lambda$, yet it also accurately predicts thermal emission from a blackbody at short wavelengths.

Other than illumination, thermal radiation is also the most abundant source of energy on our planet. The total solar daytime irradiance reaches 1,360 W/m². This power density suffices to meet the world's energy needs for an entire year, if one collected *all* the energy that reaches the earth for just one hour. Harvesting this abundant renewable energy source for electricity is the subject of a large portion of modern research[4], with solar photovoltaic cells having reached the stage of a mature technology[5,6]. Solar photovoltaic cells harness the portion of solar photons with frequencies in the visible range. For a single junction photovoltaic cell, the fundamental efficiency limit of roughly 30% was derived by Shockley and Queisser in 1961[7].

Together with the sun, terrestrial objects in the macro- as well as in the micro-scale emit mid-IR radiation as a result of local heating. Examples range from exhaust gases from a power plant, to a hot stove top, to an operating microprocessor. Due to its abundance, this radiant heat, often termed *waste heat*, presents a significant opportunity for heat-to-electricity energy conversion, recycling, and storage. A significant portion of waste heat resides at temperatures below 1000 K[8], and this, from a thermodynamic point of view, hinders its efficient conversion. Since the ultimate thermodynamic efficiency of a heat engine is $1 - T_C/T_H$ (the Carnot limit), where $T_C$ and $T_H$ are the temperatures of the cold and hot reservoirs, respectively, reaching high efficiencies at low-grade waste heat temperatures ($T_H < 1000$ K) is a challenge[9].

a)georgia.papadakis@icfo.eu

To harness waste heat, one can use a *thermo-photovoltaic* system, which operates similarly to a solar photovoltaic cell. In the thermo-photovoltaic case, however, instead of the sun, heat elevates the temperature of a local thermal emitter, which in turn emits radiation towards a low-band gap photovoltaic cell, whose band gap energy corresponds to the average energy of photons from the thermal emitter[10]. The characteristics of thermal emission can be engineered to match the properties of the corresponding photovoltaic cell, thus, theoretical predictions estimate thermo-photovoltaics performance near thermodynamic limits[5,11–13]. Experimental demonstrations of thermo-photovoltaic systems with efficiencies in the range of 40% have already been reported, although this peak efficiency was reached at high emitter temperatures (2400°C)[5].

Beyond energy and lighting, harnessing and controlling thermal radiation finds applications in any technological platform where regulating a temperature is relevant. One notable example is radiative cooling[14–23]. Radiative cooling is a technique via which the particular features of the spectrum of the atmosphere's transmittance are exploited to decrease the temperature of objects on earth[24]. In particular, the atmosphere is transparent between 8 and 13 $\mu m$, a frequency range which corresponds to the peak of the emission of bodies at room temperature. Therefore, all bodies at room temperature emit thermal radiation which can escape the atmosphere and thermalize with the coldness of the universe at 3K. In order to cool down, however, the body's total energy balance needs to be negative: the energy the body absorbs from its surroundings must be smaller than the energy it emits[25]. For this reason, radiative cooling devices have been designed using a nearly ideal mirror to reject solar photons from the sun, while at the same time, they emit efficiently at frequencies within the atmospheric transparency window.[26–28].

Other applications in which the control of thermal radiation is crucial include thermal camouflaging[29–33] and circuitry[34–43]. Applications in sensing[44], thermal imaging[45,46], and spectroscopy[47,48] are also relevant to thermal photonics. The field of thermal photonics aims to tailor the incoherent thermal emission from blackbodies to meet the requirements of various applications. Very interesting results demonstrating a high degree of both spatial and temporal coherence have been recently reported[13].

This tutorial is structured as follows: first, we introduce our notation and theoretical framework, and distinguish between the thermal near-field and far-field. We briefly discuss the spectral and spatial coherence of both near- and far- field thermal radiation, and comment on the techniques adopted to measure thermal emission in both ranges. Next, we describe prominent mechanisms available for dynamic control of thermal emission and their potential applications. We focus individually on each distinct phenomenon which can be used to achieve dynamic modulation, yet we note that, in practice, sev-eral of the described phenomena may be combined to design mid-IR photonic devices for applications.

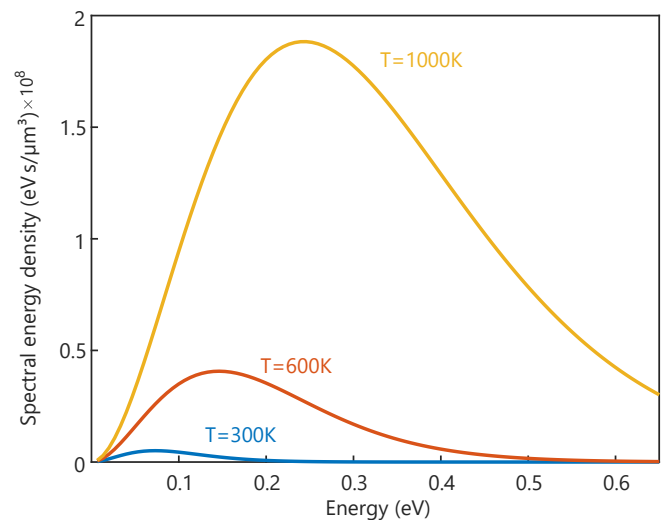

FIG. 1. Spectral energy density for a blackbody calculated using Eq. 1 for T= 300 K, 600 K and 1000K.

## II. PLANCK'S LAW OF THERMAL RADIATION

Planck's law describes how the temperature of a body determines the amplitude and spectrum of its thermal emission[1]. It states that the spectral energy density (energy per unit volume), $\phi(\omega)$, for an emitter at temperature $T$ and frequency $\omega$ can be expressed as:

$$\phi(\omega) = E(\omega, T) \times \text{DOS}, \qquad (1)$$

where $E(\omega, T)$ is the mean thermal energy per photon, given by $E(\omega, T) = \hbar \omega N(\omega, T)$, with $\hbar$ being the reduced Planck's constant, and $N(\omega, T)$, the photon number, which is determined via the Bose-Einstein distribution:

$$N = \frac{1}{e^{\frac{\hbar \omega}{k_B T}} - 1}, \qquad (2)$$

where $k_B$ is the Boltzmann constant. The factor DOS in Eq. 1 is the photonic density of states, which quantifies the number of photonic states available per frequency and unit volume.

Via Eq. 1, it already becomes possible to identify a mechanism available for dynamic tuning of thermal emission: modulation of the temperature of a blackbody emitter. As shown in Fig. 1, changing the temperature in Eq. 1 leads to significant changes in both the central frequency and the amplitude of the spectral density of a blackbody emitter. Tuning the temperature of an emitter is what enables thermo-optical modulation, as well as a mechanism to induce transitions in phase change materials, as we will discuss later on.

Planck's law assumes that the distance from an emitting body, at which the spectral flux is measured, as well as the dimensions of the body itself, are larger than the thermal wavelength, which is given by:[49]

$$\lambda_T = \pi^{2/3} \frac{\hbar c}{k_B T}. \tag{3}$$

As we shall see below, raising these assumptions and considering separation distances much smaller than $\lambda_T$ yields thermal emission and radiative heat transfer rates that can exceed the blackbody limit[50].

## III. FAR-FIELD AND NEAR-FIELD RADIATIVE HEAT TRANSFER

With respect to $\lambda_T$ as defined in the previous section, we broadly distinguish between two spatial regions: the region that is sufficiently far away from a thermal emitter, for which $d \gg \lambda_T$, or far-field (FF), and the region in close proximity to the emitter for which $d \ll \lambda_T$, termed thermal near-field (NF). We note that an alternative definition to NF that takes into account the emitter size and is irrespective of its temperature is also discussed in literature[51,52].

In order to understand the key differences between these two regions, let us start by considering a local monochromatic emitter in an otherwise lossless, isotropic and homogeneous background, placed at $r = r_0$. The electric field at a point $r$ can be described as a plane wave:

$$E(r, \omega) = E_0 \, e^{i[k \cdot (r - r_0) - \omega t]}, \tag{4}$$

where $E_0$ is the amplitude of the field at $r_0$, $k$ is the wavevector, $\omega$ is the angular frequency and $t$ is time.

If the wavevector $k$ is a real number, then at a distance r from the source, the field acquires a phase $e^{ik \cdot (r - r_0)}$ while its amplitude is preserved. We shall refer to light with such a wavevector as a *propagating* wave. By definition, FF radiation is entirely comprised of propagating waves. In Fourier space, as shown in Fig. 2(c), propagating waves are confined to the volume inside the lightcone defined by $k_0 = n\frac{\omega}{c}$, $n$ being the refractive index of the surrounding medium and $c$ the speed of light in vacuum[51]. By contrast, if the transverse component of the wavevector extends to values of $k$ outside the lightcone, the wavevector in Eq. 4 becomes complex, resulting in an exponentially decaying wave away from its source, also termed *evanescent wave*.

The scenario of two objects exchanging radiative heat in the FF is shown in Fig. 2 (a) for the case of planar surfaces at different temperatures. By contrast, when the objects are placed in close proximity ($d \ll \lambda_T$), we shall refer to them as being in the NF of one another, Fig. 2 (b), where interesting phenomena beyond Planck's law can emerge[53].

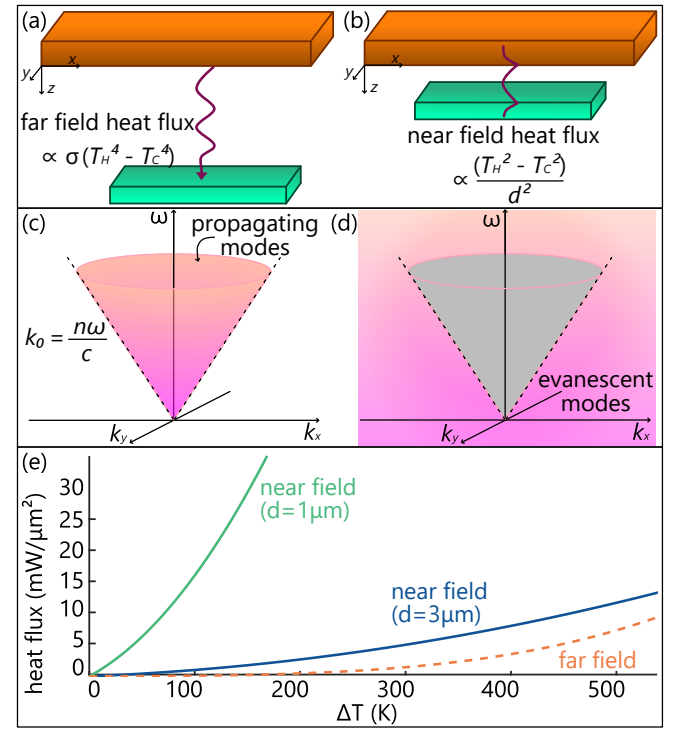

FIG. 2. Schematic depiction of two planar surfaces exchanging thermal radiation in (a) the far-field (FF) and (b) near-field (NF). FF, propagating waves preserve their amplitudes with their phase oscillating away from the planar surfaces, while the amplitude of evanescent waves decays exponentially. (c, d) Schematic depiction of FF and NF waves in Fourier space. For propagating waves, the amplitude of the wavevector ($k$) is bounded (c) inside the light cone, delimited by $k_0 = n\frac{\omega}{c}$ and denoted by dashed lines, whereas (d) evanescent waves lie *outside* the lightcone. In (e), the NF heat flux calculated between two SiC planar surfaces with $0 < \omega < 10\,\Omega_{SiC}$, $\Omega_{SiC}$ being the resonant frequency of bulk SiC, and $k_0 < k_t < \frac{1}{d}$ at different distances is shown in comparison to the FF, as a function of the temperature difference $\Delta T = T_H - T_C$.

### A. Far-field

In the FF, we can express the DOS in Eq. 1 as:

$$DOS_{FF} = \frac{\omega^2}{\pi^2 c^3}. \tag{5}$$

Plugging this in Planck's law (Eq. 1), we can evaluate the electromagnetic energy emitted from a blackbody at temperature $T$. It is useful to introduce now a quantity called *emissivity*, which serves as a metric of the efficiency of an emitting body as a thermal emitter[54]. It is defined as the ratio between the energy emitted by a body at temperature $T$ and that emitted by a blackbody at the same temperature. Hence, blackbodies are characterised by unitary emissivity, i.e.: $\mathcal{E}_{BB} = 1$, at all frequencies. The emissivity, $\mathcal{E}$, obtains values between 0 (a perfect reflector) and 1 (a perfect emitter). We note that the emissivity, as defined here, is a FF property of emitters with no counterpart in the NF. Due to Kirchhoff's law of thermal radiation[55], by reciprocity, $\mathcal{E}$ is also equal to the optical absorption $\mathcal{A}$ of a material. Often, in literature, a more realistic emitter is considered, termed *greybody*, characterized by a constant emissivity $\mathcal{E} < \mathcal{E}_{BB}$. This is still an idealization as emissivity in general depends on frequency, but can be a good approximation for some materials in given spectral ranges. As an example, in Fig. 3, the spectral energy density of a greybody with $\mathcal{E} = 0.25 \mathcal{E}_{BB}$ (orange curve) is compared to that of a blackbody (red curve).

The DOS$_{FF}$ of equation 5 is the density of states evaluated in vacuum. If, instead of vacuum, the emitter is embedded into a lossless medium, the intensity of the radiation scales proportionally to $n^2 = \varepsilon$, where $n$ is the refractive index of that medium and $\varepsilon$ is its dielectric function[56]. This still holds if $n$ is both a function of frequency and position. The dielectric function of materials is a key property for actively tailoring thermal emission, as will be discussed below. In the most general case, $\varepsilon$ is a $3 \times 3$ tensor with up to 6 independent degrees of freedom. For isotropic materials $\varepsilon$ is a scalar quantity. In the following we will see examples of dynamic control of the scalar dielectric function to modulate thermal emission, such as thermo-optical and electrostatic modulations, as well as mechanisms in which the tensor nature of $\varepsilon$ is fundamental, as is the case for magneto-optical modulation.

Let us consider the FF heat exchange between two greybodies as in Fig. 2 (a). The heat transfer between them can be accurately described by the Stefan-Boltzmann law. Upon integrating spatially and spectrally the heat flux exchange of two bodies, one at temperature $T_C$ and the other at temperature $T_H$, for $T_H > T_C$, the Stefan-Boltzmann law predicts a total heat exchange that is proportional to $(T_H^4 - T_C^4)$. We therefore note that, in the FF, thermal emission as well as the total radiative heat transfer is independent of the distance from the emitter.

On the other hand, when the heat exchange between the two bodies occurs at distances *comparable* or *below* the thermal wavelength $\lambda_T$, as in Fig. 2 (b), the modes that dominantly contribute to the heat transfer do not follow Stefan-Boltzmann law (blue and green curves), as can be seen in Fig. 2(e). This is explained in the following section.

## B. Near-field

In close proximity to the emitter, the radiative heat flux is mediated by evanescent fields that are highly localized close to the emitting body, and quickly decay away from it, as shown for example in Fig. 2 (b) for the case of two planar bodies. In the NF, the DOS assumes different values depending on how far from the emitter it is measured, hence it is usually referred to as a *local* density of states, LDOS, to take into account this spatial depen-dence. The LDOS can be calculated by considering a thermal emitter as a superposition of point dipoles and using the dyadic Green's function to estimate the electric and magnetic fields generated by their oscillation[52]. To relate these oscillations to macroscopic quantities (for instance, the temperature), we impose the fluctuation-dissipation theorem on the current density generated by these dipoles. This determines its statistical correlation. We can therefore write the LDOS for a thermal emitter[50,57,58]. Following Joulain *et al.*,[59] we highlight the main steps towards the derivation of the LDOS, albeit referring the reader to dedicated resources for further details[59–61]. The electric field correlation function can be written as:

$$\mathscr{E}_{ij}(\mathbf{r}, \mathbf{r}', t - t') = \frac{1}{2\pi} \int_{-\infty}^{\infty} d\omega\, \mathscr{E}_{ij}(\mathbf{r}, \mathbf{r}', \omega)\, e^{i\omega(t-t')} =$$
$$= \langle E_i(\mathbf{r}, t) E_j^*(\mathbf{r}', t') \rangle .$$

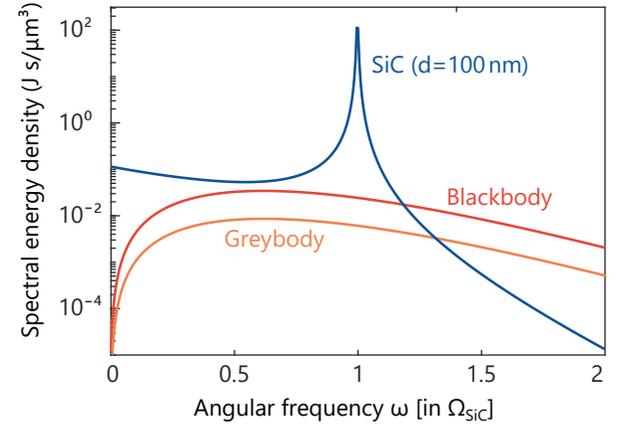

FIG. 3. Spectral energy density calculated using equation 1 (we note that the y axis is in logarithmic scale). In blue, the spectral energy density is calculated in the NF of a SiC emitter at a distance $d = 100\,nm$. We see that the NF thermal emission shows a highly enhanced energy density around a narrowband peak. In red and orange, we show the FF spectral energy densities evaluated for a black ($\mathcal{E}_{BB} = 1$) and a greybody ($\mathcal{E} = 0.25$), respectively. The $x$ axis is normalized with respect to $\Omega_{SiC} = \sqrt{\frac{\varepsilon_\infty \omega_{LO}^2 + \omega_{TO}^2}{1 + \varepsilon_\infty}}$, which is the phonon polariton resonance frequency for bulk SiC.

The electric field can be written from the electric current density as:

$$\mathbf{E}(\mathbf{r}, \omega) = i\mu_0 \omega \int \mathbf{G}^E(\mathbf{r}, \mathbf{r}', \omega) \cdot \mathbf{j}(\mathbf{r}') d^3 \mathbf{r}',$$

where $\mathbf{G}^E$ is the electric field's dyadic Green's function, $\mathbf{j}$ is the current density and $\mu_0$ is the magnetic vacuum permeability[51]. Via fluctuation-dissipation theorem, we

can write:

$$\mathscr{E}_{ij}(\mathbf{r}, \mathbf{r}', \omega) = \frac{\hbar\omega}{e^{\frac{\hbar\omega}{k_B T}} - 1} \frac{\mu_0\,\omega}{2\pi} \operatorname{Im}[G_{ij}^E(\mathbf{r}, \mathbf{r}', \omega)].$$

A similar procedure can be applied to the magnetic field, using its field's correlation and dyadic Green's function. Adding the electric and magnetic contributions, and considering only the positive frequencies, it is possible to obtain the local spectral energy density:

$$\phi(\mathbf{r}, \omega) = \frac{\hbar\omega}{e^{\frac{\hbar\omega}{k_B T}} - 1} \frac{\omega}{\pi c^2} \operatorname{Im}\{\operatorname{Tr}[\mathbf{G}^E(\mathbf{r}, \mathbf{r}', \omega) + \mathbf{G}^H(\mathbf{r}, \mathbf{r}', \omega)]\},$$

from which the local density of states can be written as:

$$\text{LDOS}_{\text{NF}} = \frac{\omega}{\pi c^2} \operatorname{Im}\{\operatorname{Tr}[\mathbf{G}^E(\mathbf{r}, \mathbf{r}', \omega) + \mathbf{G}^H(\mathbf{r}, \mathbf{r}', \omega)]\}. \quad (6)$$

The Green's functions in Eq. 6 depend on the considered geometry. Importantly, in vacuum, $\operatorname{Im}[\operatorname{Tr}(\mathbf{G}^E)] = \operatorname{Im}[\operatorname{Tr}(\mathbf{G}^H)]$. From the boundary conditions at infinity $r \to \infty$, one retrieves the $\text{DOS}_{\text{FF}}$ which we wrote in Eq. 5. There are also geometries for which the magnetic Green's function is negligible compared to the electric one, leading to a simplified expression for the LDOS. For example, if we consider a planar emitter, calculating the fields at a distance $d \ll \lambda$, in other words in the quasistatic limit, for some materials it is possible to approximate the LDOS as[62]:

$$\text{LDOS}_{\text{NF}} = \frac{1}{4\pi^2\omega d^3} \frac{\operatorname{Im}[\varepsilon(\omega)]}{|\varepsilon(\omega) + 1|^2}, \quad (7)$$

where $\varepsilon$ is the dielectric function of the emitting material, which is a function of frequency. This approximation is useful for obtaining insight on two important concepts. Firstly, we see that in the lossless limit ($\operatorname{Im}[\epsilon(\omega)] \to 0$), the $\text{LDOS}_{\text{NF}}$ vanishes and no thermal emission occurs, as expected. This unveils clearly the tight link between optical losses and thermal radiation. Secondly, Eq. 7 is instructive to understand near field heat transfer in the presence of polaritons.

Polaritons are quasiparticles which derive from the collective oscillation of light and matter. For example, they appear in materials which present a large enough carrier density and mobility, such as metals, to allow the electrons to undergo oscillations driven by the electric field at given frequencies. In the following, unless specified, we will use the terms "metals" and "plasmonic materials" interchangeably. Analogously, in polar dielectric materials, phonon polaritons can be excited from the coupling between the electromagnetic field with the material's optical phonons[63–65]. In the following, this type of material will be referred to as "polar material". In both cases, the excitation of a polariton is dependent on the frequency of the incident light. Therefore, the response function of a polaritonic material will be resonant around a frequency, at which matter (charge or lattice vibration in plasmonic and polar media, respectively) is oscillating synchronously with the excitation, with minimal lag[66]. This response function is given by the linear polarizability of the material, which is proportional to the dielectric function[67]. The two resonant models most commonly used to describe the dielectric function of polaritonic materials are the Drude model, suitable for metals, and the Lorentz model, suitable for polar semiconductors and dielectrics[68,69]. According to the Drude model, the dielectric function of a plasmonic material can be written as:

$$\varepsilon_{pl}(\omega) = \varepsilon_\infty \left(1 - \frac{\omega_p^2}{\omega^2 - i\gamma\omega}\right), \quad (8)$$

where $\varepsilon_\infty$ is the the dielectric constant evaluated at the limit for very high frequencies, usually $\varepsilon_\infty \approx 1$ for metals, $\omega_p$ is the plasma frequency and $\gamma$ is the damping rate due to the oscillating electrons colliding with each other. The Lorentz model, on the other hand, predicts the dielectric function for a polar material[70]:

$$\varepsilon_{po}(\omega) = \varepsilon_\infty \left(1 + \frac{\omega_{\text{LO}}^2 - \omega_{\text{TO}}^2}{\omega_{\text{TO}}^2 - \omega^2 - i\gamma\omega}\right), \quad (9)$$

with $\omega_{\text{LO}}$ and $\omega_{\text{TO}}$ being the resonant frequencies corresponding to the longitudinal and transverse optical phonon resonances in the polaritonic material[71–73]. The frequencies $\omega_{\text{TO}}$ and $\omega_{\text{LO}}$ delimit the region in which the dielectric function of the material is negative, which defines the Reststrahlen band[64,74].

In Fig. 4, the real part of the dielectric function of a metal (Drude) and a polar dielectric (Lorentz) material are plotted in orange and blue, respectively.

At the resonance frequency of a polaritonic material[64,65], when $\epsilon(\omega) = -1$, the $\text{LDOS}_{\text{NF}}$ resonates. Therefore, when polaritonic excitations are available, the LDOS in the NF can dramatically exceed the DOS of the FF, leading to a spectral energy density which surpasses the blackbody spectrum in magnitude. In this case, it is often said that one can obtain "super-Planckian" thermal emission[76]. Moving away from the emitter, the magnitude of the fields decreases exponentially, and both the $\text{LDOS}_{\text{NF}}$ and the thermal emission spectrum ought to converge to their FF value, which is independent of the distance.

Where does the difference between thermal NF and FF come from? As discussed above, in contrast to the FF, waves that occur in the NF have an in-plane wavevector component $k_t = \sqrt{k_x^2 + k_y^2}$ that lies *outside* the vacuum lightcone as shown in Fig. 2(d), in other words $k_t > k_0$. Since the total wavevector satisfies $k = |\mathbf{k}| = \sqrt{k_t^2 + k_z^2}$, this requires $k_z \in \mathcal{I}$. This explains the exponential decay of the evanescent excitation in the direction orthogonal to $k_t$ in vacuum, which we can refer to as $z$ without loss of generality (following the reference system of Fig. 2 (a, b)). Hence, along $z$, the field does not propagate, but rather decays away from the source as $e^{-k_z z}$. In close proximity to the emitting surface, this localization of the electromagnetic field justifies strong thermal emission, which explains the term "super-Planckian". Such an enhancement is due to the addition of either frustrated modes in any material or evanescent modes in plasmonic and polar media with a negative permittivity[77]. In the latter case, the evanescent modes are typically resonant, therefore the NF enhancement of heat transfer occurs resonantly[78]. This is shown in Fig. 3, where we plot the NF spectral energy density calculated at $d = 100$ nm away from a SiC semi-infinite planar emitter at $T = 300$ K. From this curve, it is seen that the energy per unit volume above SiC surpasses that of a blackbody ($\mathcal{E} = 1$, red curve) and a greybody ($\mathcal{E} = 0.25$, orange curve). This occurs, as we said, in the Reststrahlen band of SiC where a surface phonon polariton emerges[65,79].

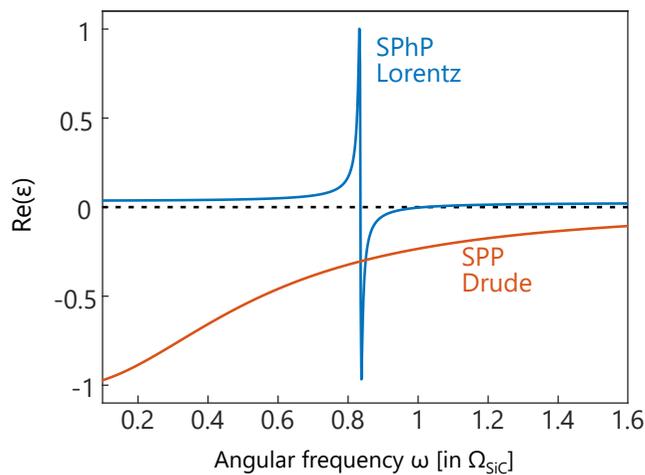

FIG. 4. Typical trend of the real part of the dielectric function of a metal (orange) and a polar dielectric (blue). Both curves have been independently normalized to their maxima (minima). The dielectric function of the metal, described via Drude model[68,75], is a damped oscillator resonant about the frequency of excitation of surface plasmons. Polar dielectrics, instead, follow Lorentz model, in which two resonant frequencies exist, corresponding to the longitudinal and transverse optical phonons[68,73]. Polaritonic surface excitations are excellent carriers of thermal radiation in the NF[53].

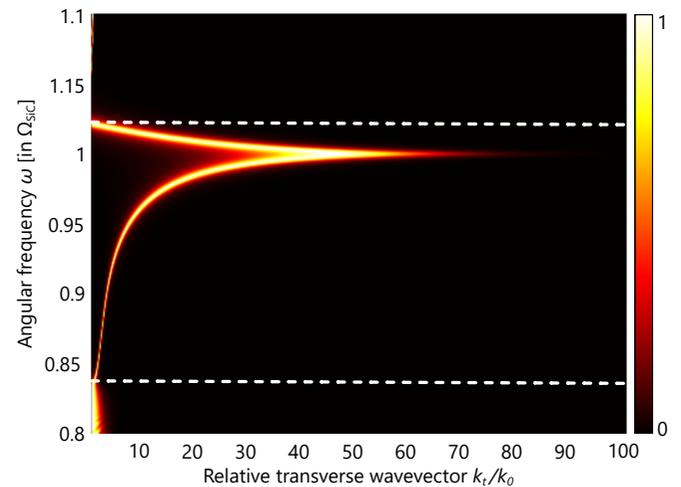

FIG. 5. Photon transmission probability between two semi-infinite layers of SiC separated by a 100 nm vacuum gap. The $y$ axis is normalized with respect to $\Omega_{SiC} = \sqrt{\frac{\varepsilon_{\infty}\omega_{LO}^2 + \omega_{TO}^2}{1+\varepsilon_{\infty}}}$, which is the phonon polariton resonance frequency for bulk SiC. The two dashed lines correspond to $\omega = \omega_{LO}$ and $\omega = \omega_{TO}$ and enclose the Reststrahlen band.

Just as we did in the case of FF [Fig. 2 (a)], let us also consider the case of two planar surfaces, in this case exchanging thermal radiation in the NF, as shown in Fig. 2 (b). In this case, as long as there is sufficient spatial overlap between the polaritonic waves on the two respective interfaces, the total, frequency-integrated heat exchange can surpass the Stefan-Boltzmann law. The photon transmission probability $\xi$, the probability that a photon emitted by body 1 is absorbed by body 2, needs to be integrated across all available wavevectors, to take into account the prominent role of surface modes at very small distances. In figure 5 we plot this quantity ($\xi$) for two planar semi-infinite SiC planes separated by 100nm.

From figure 5, the resonant behavior of the $LDOS_{NF}$ is evident, as the transmission probability is close to unity in the frequency range where SiC is resonant, and close to zero everywhere else[65,80]. When discussing electrostatic gating, we will see an example (See Fig. 10) of how the dynamic tuning of $\xi$, modulates NF heat transfer.

By considering heat transfer between planar structures and assuming that material properties vary slowly with frequency, it is possible to approximate the total integrated near field heat exchange as proportional to $\frac{T_H^2 - T_C^2}{d^2}$, where $d$ is the distance from the source[57,79–81]. This is unlike the FF case where the dependence was proportional to $\approx T^4$. In Fig 2(e), we demonstrate this difference. The dashed red orange curve pertains to the FF total heat exchange between two planar surfaces, for $d > \lambda_T$. By contrast, the blue and green curves correspond to calculations in the NF, for $d = 3\mu m$ and $1\mu m$, respectively calculated for SiC around its resonance frequency. It is seen that the NF heat flux exceeds the FF one. Tuning the distance between the two bodies exchanging heat radiatively in the NF is yet another method to modulate thermal radiation.

## IV. SPATIAL AND TEMPORAL COHERENCE OF THERMAL EMISSION

Thermal emission measured in the NF is typically a narrowband phenomenon[12]. Conversely, the black or greybody spectra evaluated in the FF are generally broadband. Even for materials with strong mid-IR resonances, such as SiC depicted in Fig. 3 and other polaritonic ones[64], their thermal emission spectrum becomes broadband and approaches the blackbody spectrum in the FF.

Other than temporal coherence, we can also consider the spatial coherence of NF and FF thermal radiation. Thermal emission is a statistical process which follows the Bose-Einstein distribution, and thermal currents satisfy the fluctuation-dissipation theorem. Through this, the degree of spatial coherence can be estimated evaluating the spatial correlation function between the field at any two separate points[82]. This corresponds to answering the question *"How much is the value of the electromagnetic field at point A influenced by the value of the field at point B?"*. Performing this analysis, the FF correlation length is estimated to be approximately $\lambda/2$ for any emitter[82].

In the NF, on the other hand, the correlation length is material-dependent. For materials that are transparent in the thermal emission wavelength range, the NF correlation length is equal to the FF one[82], as no emission occurs inside the medium. On the other hand, for absorbing media, if the penetration depth is *smaller* than the wavelength, then the spatial correlation length becomes comparable to the penetration depth, and can be as small as $0.06\,\lambda$[82]. Nevertheless, for polaritonic media, e.g. media that support surface plasmon polaritons or surface phonon polaritons, the correlation length can be significantly larger than this, at frequencies for which these modes exist, reaching values as large as $65\,\lambda$[82]. Polaritonic surface waves are highly correlated, hence radiation mediated by a surface mode inherits its spatial correlation characteristics even when thermally generated.

Being a collective excitation of light and matter, surface polaritons exist near the material's interface with another medium and are, therefore, a NF characteristic. Using nanostructures such as cavities and gratings[83], however, their coherence can be extended to the FF, by diffracting surface modes and thereby achieving temporally coherent and directional FF thermal radiation[84].

## V. MEASURING THERMAL EMISSION

A good thermal emitter, such as a black or greybody, must also be a good absorber. This arises from Kirchhoff's law of thermal radiation[55], which states that the emissivity $\mathcal{E}$ and absorptivity $\mathcal{A}$ must be equal, per frequency, polarization, and angle of incidence[55]. Indeed, in a reciprocal material, due to time reversal symmetry, anything other than $\mathcal{E}(\omega, \theta) = \mathcal{A}(\omega, \theta)$ would violate the second law of thermodynamics[54,85].

Due to Kirchhoff's law, in reciprocal materials, measuring thermal emissivity and absorptivity ought to give the same result. This facilitates experiments, as measuring thermal emission requires heating up the sample to high temperatures, and collecting the emitted light at various angles, which can have low intensity and signal to noise ratio. This is especially challenging at mid-IR frequencies, where water vapor in the atmosphere has significant absorption, which further reduces detectability. Direct measurements of thermal emissivity have been performed in some seminal works[83,84,86–92].

In most reported results, however, thermal emission is derived indirectly, via absorptivity measurements, which are typically straightforward. From energy conservation, the incident light onto a sample, having an amplitude of unity for the sake of argument, should be either absorbed ($\mathcal{A}$), reflected ($\mathcal{R}$), or transmitted ($\mathcal{T}$), in other words $1 = \mathcal{A} + \mathcal{R} + \mathcal{T}$. Thus, for opaque samples or samples deposited on a reflector, for which $\mathcal{T} = 0$, thermal emissivity is determined via a measurement of absorption, which in turn equals $\mathcal{A} = 1 - \mathcal{R}$. Thus, the emissivity is obtained via $\mathcal{E} = 1 - \mathcal{R}$.

As discussed above, thermal emission in the FF is mediated by propagating modes (Fig. 2 (a)), whereas in the NF it is mediated by evanescent ones (Fig. 2 (b)). Thus, measuring thermal emission in the FF probes propagating modes, whereas in the NF it requires probing evanescent excitations which vanish rapidly away from the emitter/absorber[93]. Typically, FF thermal emission experiments involve a Fourier Transform IR (FTIR) spectrometer that measures the emitted spectrum either in transmission or reflection[84,86]. In contrast, NF thermal emission measurements are non-standard and more challenging.

First, depending on the geometry, e.g. plane-to-plane[94,95], tip-to-plane[96–99] or other, the scaling law of thermal emission as a function of the distance from the emitting body differs significantly[79]. In addition, since NF radiative heat transfer is orders of magnitude larger than its FF counterpart[53,100], when attempting to measure thermal emission in the NF, inevitably, the emitting sample transfers considerable heat to any probe used for collection, thus elevating its temperature, which influences the spectral profile measured. Moreover, since NF thermal emission is mediated via evanescent fields that lie outside the lightcone (Fig. 2 (d)), as we saw earlier, these do not couple to the FF by default. There are important experimental examples in which thermally emitted NFs have been scanned locally using sharp tips[47,99,101–104]. Nevertheless, electromagnetic fields scattered and collected via a tip involve *all* available wavenumbers without any means of spatially resolving them into different $\mathbf{k}$ channels. In contrast to FF thermal emission measurements where one can easily obtain a thermal emission spectrum, in the NF, typically, one measures the spatially and spectrally integrated heat, via calorimetry[105,106]. Novel techniques aimed at experimentally resolving a thermal emission spectrum in the near-field are current research efforts[47,97,102].

## VI. DYNAMICAL TUNING MECHANISMS

Static engineering of thermal radiation involves designing a thermal emitter with specific radiative properties[107]. This has been achieved with a vari-ety of methods, examples including the use of gratings, photonic crystals, material doping, nanostructuring and cavities[108–110]. Remarkably, by means of gratings[84,111–113] and multilayers[109], highly coherent FF emission has been obtained. Via photonic crystals, optimal designs for daytime radiative cooling have been achieved[21,114], where the emissivity is engineered to be high only within the spectral range where our planet's atmosphere is transparent. Furthermore, via nanostructuring and metamaterials, both the direction[115,116] and bandwidth of thermal emission can be tailored[117–119]. The aforementioned mechanisms have been developed to statically engineer thermal emission to fit the requirements of specific applications. Dynamic modulation, on the other hand, requires the capability of *active* tuning in time. Here, we will review some mechanisms that make this possible and highlight some recent theoretical and experimental advances.

### A. Phase change materials

A common mechanism for dynamic modulation of thermal emission involves the use of phase change materials[120–122]. These materials possess two or more structurally distinct and often reversible phases, such as amorphous and crystalline. The phase transition can be triggered by reaching a critical temperature $T_C$[123–125], above which the material's response to mid-IR radiation changes significantly.

The phase transition is volatile, when the material returns to its initial phase once the temperature is reduced to below $T_C$. In contrast, the phase transition can be non-volatile, for example when a material is triggered by a fast delivery of energy in the form of a light pulse that locally heats up the material. After the pulse, the material stays in the phase it has acquired independently of the external temperature, and can revert back to the initial phase via another *different* pulse or series of pulses[126–128]. The material behaves very differently in the two phases, hence the emission spectra possess distinctively different spectral features in these two phases. Fig. 6 is a pictorial depiction of the emissivity of a phase change material at two different phases. The two curves represent the change in emissivity that the material experiences upon the phase transition, for example, from the crystalline to the amorphous phase.

There exist two important classes of phase change materials which are commonly used in the mid-IR spectral range. The first is comprised of materials that transition between an amorphous and a crystalline phase, both dielectric ones yet with different refractive indices. A material as such is GST, a class of composites of Ge, Sb an Te with varying stoichiometry[127,130–137]. As an example, in Fig. 7 we report a result by Li *et al.*[129] in which non-volatile phase transitions have been induced in a GST sample. In panel (a) a schematic of the experiment is depicted. The areas depicted as "Written" in panel (a), have transitioned from the crystalline to the amorphous phase. The different phase between the "Written" areas and the crystalline substrate is clearly visible in the two optical measurements of panels (b) and (c), performed at different wavelengths. On the other hand, the areas indicated by "Erased" in panel (a) have undergone a complete phase cycle, from crystalline to amorphous and back to crystalline. These areas are barely distinguishable from the substrate in panels (b) and (c), highlighting the reversibility of the process.

The second class pertains to materials that transition between a dielectric and metallic phase, as it is the case for vanadium dioxide $(VO_2)$[138–144]. The metal-insulator transition of $VO_2$ occurs near $340\,K$[145]. This near-room temperature $T_C$ makes $VO_2$ highly desirable for non-volatile phase transition applications. At mid-IR frequencies, the dielectric permittivity of $VO_2$ changes dramatically upon the phase transition, from a large and positive value in the dielectric phase to a large negative value in the metallic phase[146]. This makes $VO_2$ a suitable material for efficient tunable thermal emitters, as a small temperature gradient can yield a large emissivity change. This has led to various experimental demonstrations of the effect in the FF[125,146–148]. In the NF, reports predict interesting effects such as a near-field thermal transistor[36], and $VO_2$-based nanoscale thermal diodes have been experimentally demonstrated[38].

Applications of phase change materials-based thermal emission extend to optical switching[128,129,149,150], lensing[151], engineered absorption[137,152,153] and thermal rectification[38,140,154–159].

### B. Thermo-optical modulation

Even in the absence of a crystallographic phase transition, various materials change their optical properties following a change in temperature, due to the thermo-optical effect. In this case, in contrast to phase change materials, a continuous change in temperature yields a continuous change in optical response. Such materials can be exploited to dynamically modulate thermal radiation, with examples including polymers[160] such as poly(methyl methacrylate) (PMMA)[161,162] as well as intrinsic semiconductors[163–165].

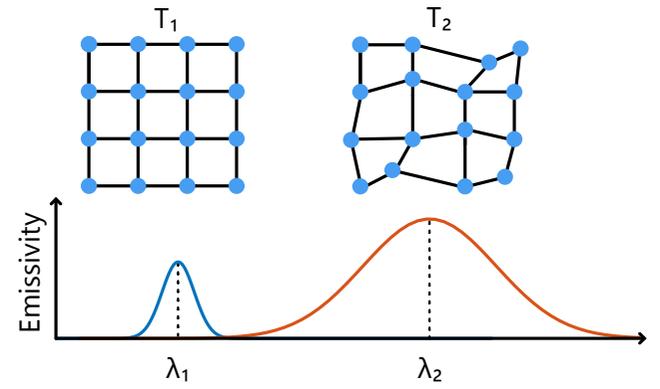

FIG. 6. Schematic depiction of phase change-based thermal emission modulation. At temperature $T_1$ the material is in its crystalline phase while at temperature $T_2$ it is amorphous. Accordingly, the emission spectrum at the two temperatures is going to be drastically different in terms of resonance wavelength, bandwidth, and amplitude.

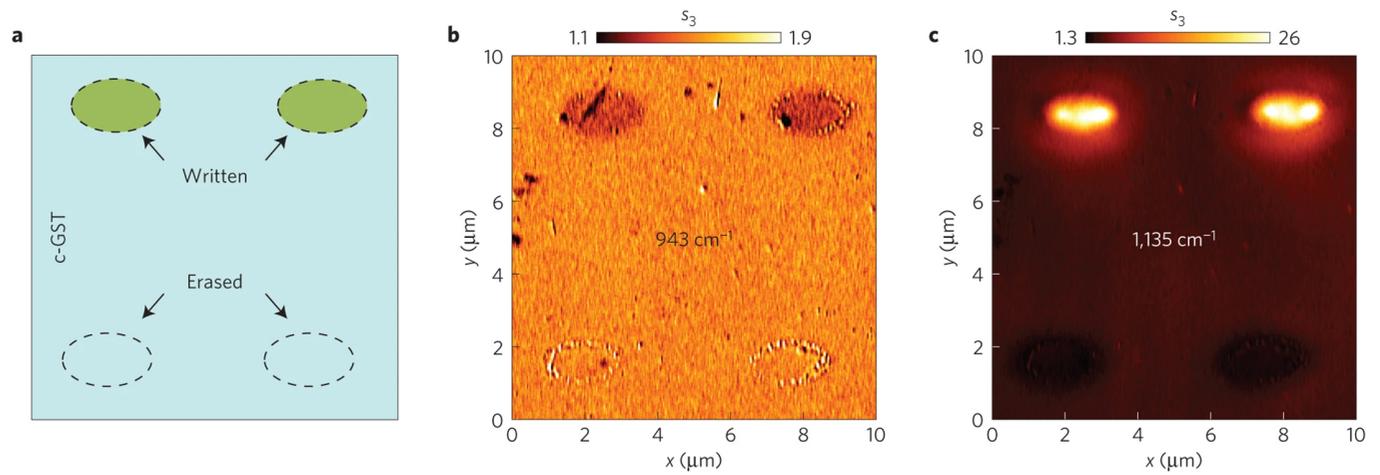

FIG. 7. Example of phase change material (GST). The four areas indicated by ellipses in panel **a** have gone through the crystalline → amorphous phase transition via 15 ns laser pulses with a power of 310 mW. Subsequently the two bottom ellipses, indicated as "Erased" in panel **a** have undergone the reverse transition from amorphous → crystalline. In **b** and **c** SNOM (Scanning Near-field Optical Microscopy) images of the sample taken outside and inside the Reststrahlen band, are shown respectively. In the Reststrahlen band surface phonon polaritons are excited in the sample. Reproduced with permission from Nature Materials 15, 870–875 (2016). Copyright Springer Nature 2016.[129]

The thermo-optical effect modifies the dielectric function of a material via temperature. For example, some polar materials change the shape and position of their material resonances at different temperatures[166,167]. This is schematically represented in Fig. 8, where the real part of a dielectric function following Lorentz model's dispersion for an idealised material is plotted corresponding to two different temperatures, with the resonance frequency shifting as the temperature is modified.

Such modulation has been used to achieve thermal rectification[166,168]. This, in turn, has led to the theoretical and experimental development of both thermal diodes[163,164,169] and transistors[36,170], fundamental devices for the control of the heat flow, for which a temperature dependent dielectric function is a requirement. In Fig. 9 an example of a thermal rectification scheme is presented, leading to a thermal diode. The working principle behind this diode is that the two materials exchanging heat have spectrally-aligned resonances at some temperature, corresponding to forward bias, yet they have mismatched resonances for the working temperature of reverse bias. In turn, this leads to a strong electromagnetic coupling with a large heat exchange in forward bias, as opposed to an inhibited heat flow in the reverse bias configuration.

While temperature modulation is slow, schemes with fast modulation can also be designed. It has been shown recently that it is possible modulate the mid-IR response of metamaterials with subwavelength elements that heat up significantly more rapidly than their environment[171,172].

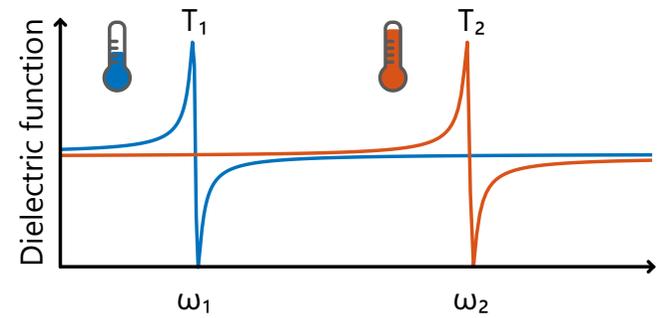

FIG. 8. Thermo-optical modulation of thermal radiation arises by changes in optical properties that a material supports at different temperatures. We depict the dielectric function of a fictitious material that has a Lorentzian lineshape. This function is resonant at frequency $\omega_1$ for a lower temperature $T_1$ (blue curve), and at a higher temperature $T_2$ (orange curve).

## C. Electrostatic gating

One of the most common mechanisms for the dynamical control of light is via electrostatic gating. Thermal radiation is no exception, and gate-tunable radiative thermal properties are being extensively investigated. In this case, one leverages the tunable carrier mobil-ity of some thermally emitting materials[173]. In particular, in materials with a high carrier mobility, electrostatic gating spatially separates the charge carriers across a p-n or metal-oxide-semiconductor (MOS) junction which, in turn, modifies the optical properties of the material. For example, electrostatic tuning of materials with free-electrons, such as semiconductors, transparent conductive oxides, semimetals etc., can be used to design field-effect thermal transistors with tunable mid-IR absorptivity[174,175]. Such an example, from[174], is shown in Fig. 10, which depicts a "thermal MOS" device. In this case, placing two identical such MOS devices across each other and separating them via a vacuum gap enables strong NF thermal modulation due to the variation of the propagation characteristics of plasmon polaritons arising from the plasmonic active regions. In Fig. 10, the photon tunnelling probability, a metric of the efficiency of NF heat transfer, is shown for two different voltages. The voltage tuning changes the propagation characteristics of surface plasmons which mediate the radiative heat transfer[174], thus modulating the NF emission of the device.

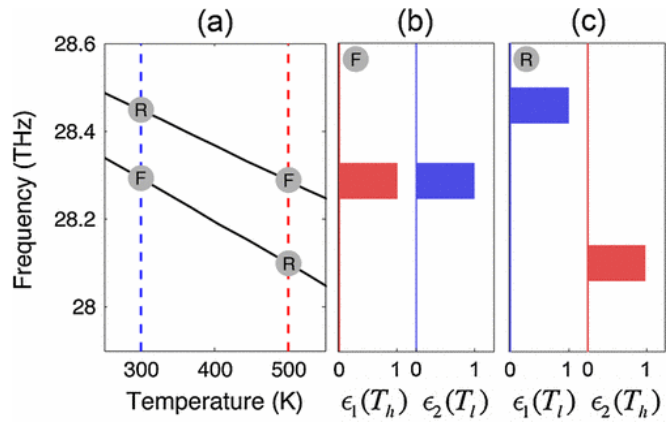

FIG. 9. Thermal rectification scheme. The black lines in panel (a) depict the temperature-dependent behaviour of the electromagnetic resonances of two materials. The circle indicates the operation points for forward (F) and reverse (R) bias. In panels (b) and (c) the two biasing configurations are shown, with the resonances being spectrally aligned for the forward bias and mismatched for the reverse bias. The spectral overlap of the two resonances in the forward bias allows for strong coupling between the two materials. On the other hand, due to the spectral separation in the reverse bias, thermal energy transfer is inhibited in this configuration. This constitutes the working principle of a thermal diode. Reproduced with permission from Phys. Rev. Lett. 104, 154301 (2010). Copyright 2010 American Physical Society[166].

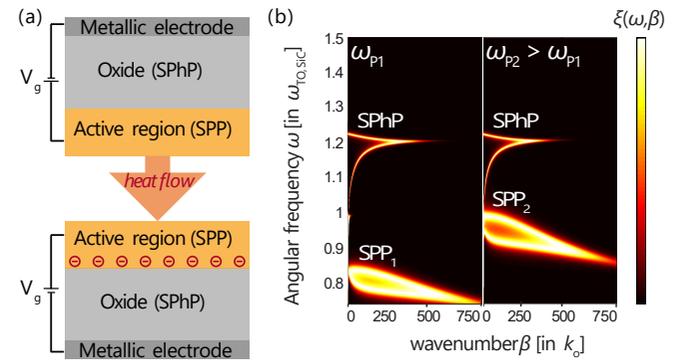

FIG. 10. Gate-tunable modulation of heat flow in a thermal MOS (Metal-Oxide-Semiconductor) device. Via the external applied voltage, the carrier density in the semiconductor layer is modified. This influences the surface plasmon it supports, which is designed to be spatially confined in the active region of the thermal MOS. When two such MOS capacitors are facing one another, as depicted in (a), the heat transfer between the two is dominated by the plasmons. This means that it can then actively be controlled via the gate bias. In (b) the photon transmission probability for two different plasma frequencies corresponding to different applied voltages. Adapted with permission from ACS Photonics 6, 709-719 (2019). Copyright 2019 American Chemical Society[174].

In ferroelectric materials, a given polarization field can be induced and reversed via an externally applied electric field[176–178]. Applying an external electric field reconfigures the ferroelectric domain walls and changes the material's thermal conductivity[179]. This allows a quick and reliable modulation of the thermal emission[178,180–182].

Similarly, electrostatic gating has been leveraged for tuning of the absorptivity, and, via reciprocity, of the emissivity, via modulation of interband energy levels in quantum well-like structures or photonic crystals[114,183,184]. Semiconductors constitute a class of materials with a dynamical and relatively fast response to gate modulation, as it is known from electronics[185]. Materials such as GaAs have shown impressive modulation of their thermal emission under electrical gating, due to the additional contribution from nonequilibrium hot electrons who can reach very high effective temperatures[186]. Another class of materials which is suited for electrostatic gating is nanostructured carbon composites, such as graphene[187] or carbon nanotubes[188], due to their high carrier mobility and thermal conductivity.

In particular, in this case, electrostatic gating results in a shift of the Fermi level of the electrons, which in turn modifies the material's optical properties[189–192]. We represent this schematically in Fig. 11, where we show two possible electronic band structures: on the left that of unbiased graphene, on the right electrically gated graphene. Un-doped graphene has a Fermi level that corresponds to the Dirac point, where the valence and conduction bands meet, termed charge neutrality point. By gating graphene, it is possible to inject charge carriers, thus promoting electrons to the conduction band and inducing a metallic behavior[193].

The doping level of graphene can also be altered via chemical means, by substitution of carbon atoms with elements from adjacent groups in the periodic table, such as B or N, as one would do with doped semiconductors[53].

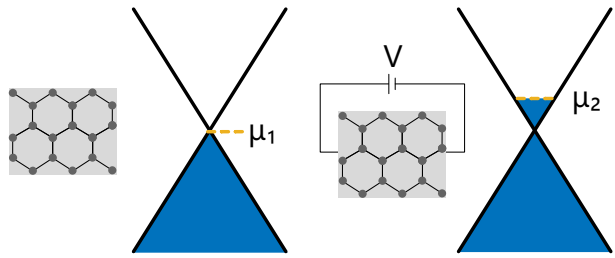

FIG. 11. Schematic representation of gate-tunability in graphene. Electronic dispersion of graphene for un-doped, not gated graphene (left), and for graphene under an external electrostatic gating (right). Modulating the applied voltage modifies the Fermi level of graphene, and, thus, its thermal emission.

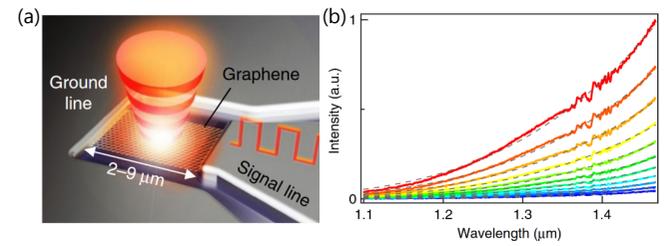

FIG. 12. Schematics of modulation of a graphene sheet on a $SiO_2/Si$ substrate via Fermi level tuning (a). The graphene is connected to source and drain electrodes (signal and ground lines). Modulated blackbody-like thermal emission is obtained from graphene by applying an input signal. The intensity of the emission from a single layer of graphene is plotted in (b), with the voltage varying from 5.5V (blue) to 8V (red). Reprinted with permission from Miyoshi et al.,, Nature Communications, 9, 1279, 2018; licensed under a Creative Commons Attribution (CC BY) license.[199]

In this case, however, the thermal radiative tuning cannot be achieved dynamically and must be coordinated with other mechanisms[174,194]. Since graphene layers are atomically thin, it is common to control the heat transfer between two or more of them by modifying their separation distance. Due to the distance dependence of the NF, particularly because of the presence of surface modes and their coupling between multi-layers, the modulation contrast can be substantial[195–198].

Fast modulation via gating graphene has also been reported. In particular, modulation rates of tens of GHz have been recently obtained in[175,199–204]. The modulation obtained via dynamic variation of the chemical potential of graphene can be exploited in tandem with other mechanisms, for example placing an intermediate graphene layer between electrostatically or thermo-optically modulated materials[205]. In Fig. 12 an example of this is shown, reprinted from reference[199], where a varying electrostatic gating on multilayer and few layer graphene results in fast modulation of its thermal emission. The emission from the graphene sample corresponds to the blackbody radiation modulated via Joule effect, with a temperature which is linearly dependent on the applied voltage. Broadband modulation with rates in the GHz range has also been experimentally observed for carbon nanotubes[206–208].

### D. Magneto-optical materials

Another method for modulating thermal radiation is via the use of magneto-optical materials. The presence of a magnetic field, whether applied externally or due to an intrinsic magnetization property, modifies the permittivity of the material[210]. As we saw earlier, the permittivity function of a material is in general a tensor quantity. Usually, we can treat it as a diagonal tensor (in uniaxial or biaxial birefringent materials), or more commonly as a scalar in isotropic materials. In the case of magneto-optical materials, however, off-diagonal terms appear in

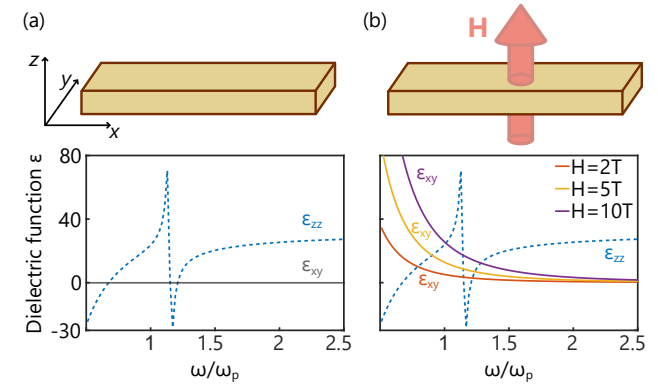

FIG. 13. Effect of magnetic field on the dielectric function of InSb – a magneto-optical material. In the absence of a magnetic field, the dielectric function is only given by the diagonal term of the permittivity tensor $\varepsilon_{zz}$. This is represented by the blue dashed curve in panel (a), while the off-diagonal term, which is equal to zero, is depicted by the black curve. As soon as the magnetic field is applied, this material exhibits off-diagonal dielectric function, as shown in panel (b). Here, the red, yellow, and purple curves correspond to the off diagonal term $\varepsilon_{xy}$ under increasing strengths of the magnetic field, from 2T to 10T, applied to the material. The material's diagonal permittivity, which is unaffected by the applied magnetic field, is again indicated via the blue dashed line. The parameters for the Lorentz model used to produce this figure are taken from Phys. Rev. B 13, 2497 (1976). Copyright 1976 American Physical Society[209].

an otherwise diagonal permittivity tensor:

$$\varepsilon = \begin{pmatrix} \varepsilon_{xx} & 0 & 0 \\ 0 & \varepsilon_{yy} & 0 \\ 0 & 0 & \varepsilon_{zz} \end{pmatrix} \rightarrow \varepsilon_{MO} = \begin{pmatrix} \varepsilon_{xx} & \boxed{\varepsilon_{xy}} & 0 \\ \boxed{\varepsilon_{yx}} & \varepsilon_{yy} & 0 \\ 0 & 0 & \varepsilon_{zz} \end{pmatrix},$$

(10)

causing the material to exhibit considerable induced optical anisotropy[210] (See Fig. 13). All well-known magneto-optical effects, such as the Faraday, Voigt, and Kerr effects, are a result of the interaction between light with a material along different coordinate directions[211–215]. Interestingly, in the presence of a magnetic field in an appropriate direction, the dielectric permittivity tensor becomes anti-symmetric in magneto-optical material systems, consequently violating Kirchhoff's law by introducing nonreciprocity[216,217]. The absorption and emission spectra of radiation are subsequently modified such that $\mathcal{E}(\theta, \omega) = \mathcal{A}(\theta, \omega)$ no longer holds true. This is relevant in various applications in mid-IR data communications and energy harvesting. For instance, it allows for one-directional propagation of light while preventing propagation in the opposite direction, which is the concept of an optical isolator[218]. A magnetic field bias can also be used to improve the efficiency in light-harvesting devices via breaking reciprocity, for example by re-directing thermal radiation from one solar PV cell to a secondary one, which increases the current of a device[219,220].

An applied magnetic field modifies the trajectories of electrons inside a metal or a semiconductor, such that a voltage transverse to the direction of the current and the applied magnetic field is developed, a phenomenon known as the Hall effect[221]. If a metallic sample is also subjected to a temperature gradient, the thermal counterpart of this effect, called Righi-Leduc effect, is observed[222]. Due to the Lorentz force acting on the electrons, the heat current by the electrons is deflected, and at the steady state, a temperature gradient, perpendicular to the initially applied temperature difference, develops[223]. This effect is not specific to electrons; it has been shown that other heat carriers in solids, such as magnons and spinons, also undergo this deflection[224–226]. Recently, photon-mediated anomalous thermal Hall effect in magnetic Weyl semimetals *without* an applied magnetic field has been reported[227]. Weyl semimetals are an emerging class of materials that have gained interest due to their distinctive topological structure. They can be thought of as three-dimensional analogues of graphene, as their conduction and valence bands overlap in energy over a certain region of the Brillouin zone[228]. These systems naturally exhibit anomalous thermal Hall effect, which allows adjusting the local temperature without changing the direction of the temperature gradient, thus, guiding the flow of heat[229]. These magnetic effects are being investigated as means for tailoring thermal emission in the FF regime[230,231].

Giant magnetoresistance, a thermo-magnetic phenomenon in which the resistivity of a material varies in the presence of a magnetic field, also holds promise for thermal radiation control at the nanoscale, among other applications. In particular, the resistivity of InSb-Ag nanoparticles has been observed to nearly double at room temperature, with applied magnetic fields as high as $2\,\mathrm{T}$[232]. However, giant magnetoresistance could be measured with weaker magnetic fields as well, by combining different magneto-optical and non-magnetic materials[232]. Furthermore, changing the direction of the applied mag-netic field allows tuning of the heat flux between two magneto-optical particles in the NF by as much as 800% for a magnetic field of 5 T[233].

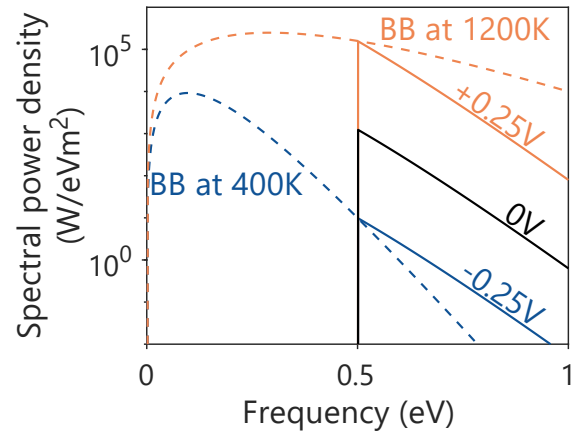

FIG. 14. Thermally emitted spectral power density from a semiconductor diode at 600K under different bias. The bandgap of the diode is 0.5 eV. The solid orange, black, and blue lines pertain to the cases where the bias voltage is 0.25, 0, and −0.25 V, respectively. At the bandgap, the spectral power density of the diode in reverse and forward bias matches that of a blackbody at 400 K and 1200 K, respectively. The data used to generate this figure are courtesy of B. Zhao (University of Houston), for details see Annual Review of Heat Transfer 23, 397-431 (2020). Copyright 2020 Begell House[236].

Modulating the magnetic field's strength is also an efficient approach for tuning radiative heat transfer in the NF. It has been reported that heat transfer in graphene gratings can be greatly modified by the presence of a magnetic field[214]. The heat transfer can further be modulated by twisting such gratings. Overall, a tuning method to implement energy modulation or multifrequency thermal communications is offered by this combined effect of magnetic field and gratings[234,235].

### E. Photon chemical potential

When a system is not in thermal equilibrium, the number of photons emitted in a given state can no longer be described by the Bose-Einstein distribution of Eq. 2. Instead, there are cases in which the emitted photons acquire a nonzero chemical potential. Let us consider, for example, a semiconductor under external bias $V$, in which the electrons are partly in the conduction and partly in the valence bands, with the electrons in each band in local equilibrium with each other. Electrons belonging to different bands have a chemical potential difference corresponding to $qV$, where $q$ is the electron charge[237]. Photons emitted, generated via the annihilation of a conduction band electron with a valence band hole, inherit the nonzero chemical potential of the material's charge carriers[236]. This influences their distribution, and the photon number statistics becomes:

$$N = \frac{1}{e^{\frac{\hbar\omega - qV}{k_B T}} - 1}. \tag{11}$$

This has a profound influence on the thermally emitted radiation, as the sign of the external bias $V$ determines either an enhancement (positive $V$) or a reduction in the spectral power density (negative $V$)[236]. This is, therefore, an indirect way to control the "apparent" temperature of the diode. For example, a diode's emission spectrum is typically narrowband (a few $k_B T$) and centered about its band gap frequency $\omega_g$. If we calculate the spectral heat flux of a diode at temperature $T$ at the energy of the band gap and equate it to the spectral heat flux of a blackbody, it is possible to define a radiant temperature $T_{\text{rad}}$[238], which can be written as:

$$T_{\text{rad}} = T\left(\frac{\hbar\omega_g}{\hbar\omega_g - qV}\right). \tag{12}$$

This is the "apparent" temperature that would be inferred by measuring the intensity of the emission, even though the actual temperature is $T$. In other words, the radiant temperature corresponds to the temperature a blackbody would have to radiate as the diode does at temperature $T$, under an external bias. Hence, photons under a positive bias seem to be emitted by an object which is hotter than $T$, while the radiation from the diode under a negative bias seems to originate from a colder object[239]. This is shown in Fig. 14, where the spectral heat flux from a photodiode under three bias conditions is plotted. The case indicated as 0V corresponds to an unbiased diode and shows the spectral power density at 600K. When a bias is applied, either forward or reverse, the spectral heat flux above the band gap is either enhanced or suppressed, respectively. Namely, at the band gap, the spectral power density can be matched to that of a blackbody at 1200K, while that of the reverse bias corresponds to a blackbody radiating at 400K.

Dynamic tuning of the chemical potential of the emitted photons can also be used to design LEDs with electrically enhanced luminescence[107,236]. Furthermore, using this mechanism, we can control the integrated total heat transfer between different photodiodes, and even revert the expected flow of heat from hot to cold, by controlling their respective external voltage[240,241].

The list of mechanisms for dynamic modulation of thermal emission we provided is certainly not exhaustive, as a wide range of other physical are being investigated that can be employed to achieve dynamic changes in thermal emission. A notable recent example pertains to mechanical strain of two-dimensional thermal emitters such as graphene or black phosphorus[242,243] or hexagonal boron nitride[244], as well as the rotation of such in-dividual monolayers[245]. With significant recent progress in nanomechanics and optomechanics, we anticipate that such approaches will also flourish as means of actively modulating mid-IR relevant radiative properties of materials.

## VII. CONCLUSIONS

We have shown a collection of mechanisms that can be exploited for dynamically modulating thermal radiation and thermal emission. Each of these mechanisms exploits the possibility of tuning an intrinsic thermal, optical, chemical or mechanical property of the material, resulting in a modification of the overall thermal emission. Engineering these mechanisms to achieve spectral features controlled and modulated in real time is relevant for various applications. The field is rapidly expanding and promises numerous more discoveries, led by breakthrough advancements in material science, nanophotonics, computing and mathematics.

## ACKNOWLEDGMENTS

The authors declare no competing financial interest. G. T. P. and K. N. N. acknowledge funding from "la Caixa" Foundation (ID 100010434) and from the European Union's Horizon 2020 research and innovation programme under the Marie Sklodowska-Curie grant agreement No 847648. The fellowship code is LCF/BQ/PI21/11830019. This work is part of the R& D project CEX2019-000910-S, funded by MCIN/AEI/10.13039/501100011033/, from Fundació Cellex, Fundació Mir-Puig, and from Generalitat de Catalunya through the CERCA program. M. F. P. acknowledges support from the Severo Ochoa Excellence Fellowship.
The authors wish to thank M. T. Enders and M. S. Saremi for their helpful comments and B. Zhao for kindly providing the data for figure 14.

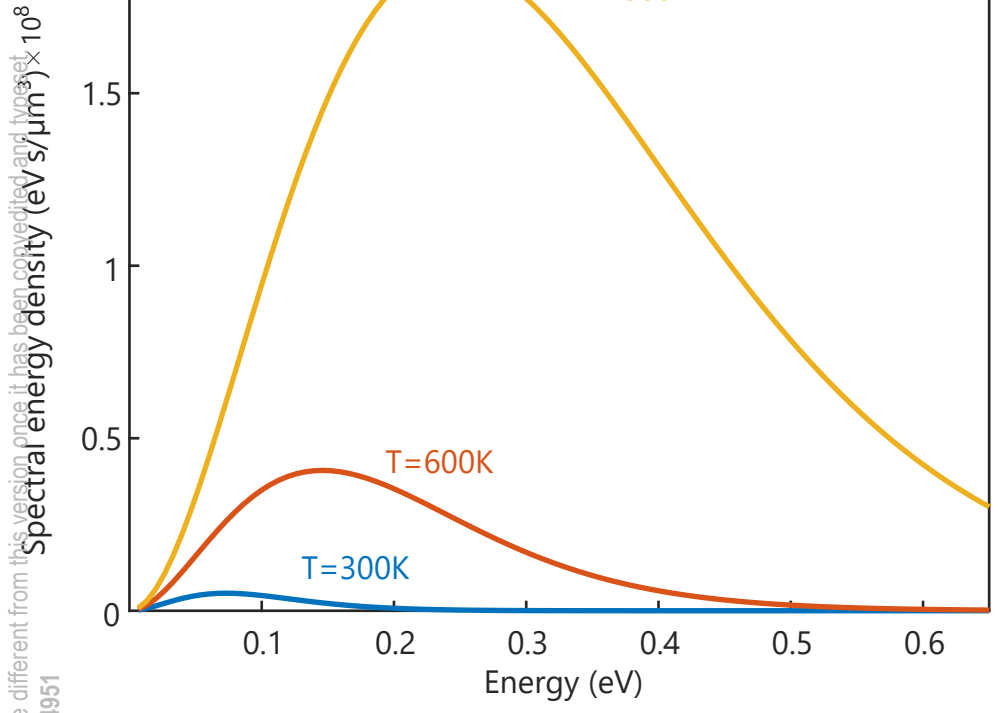

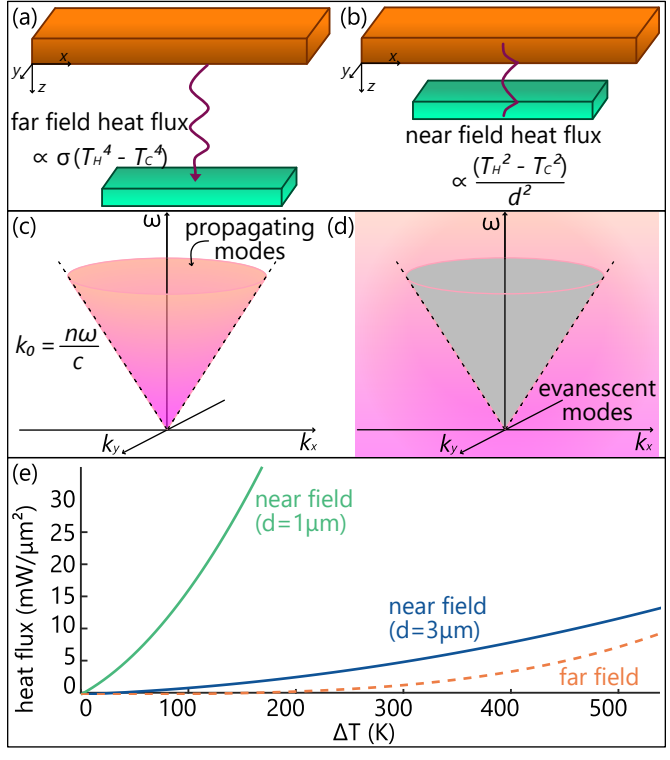

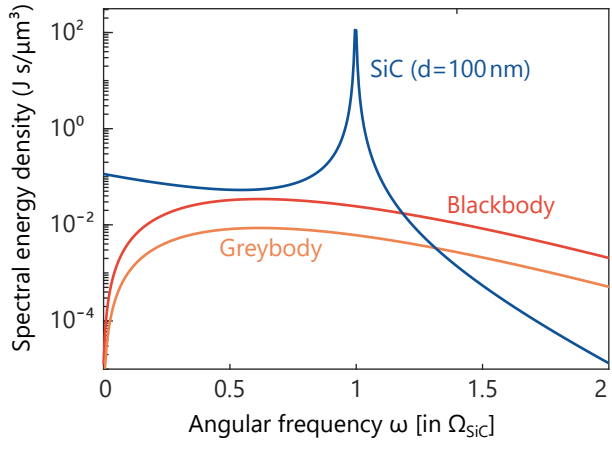

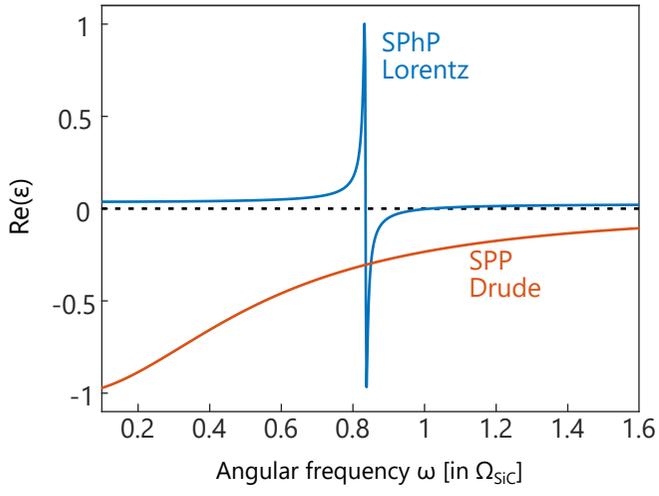

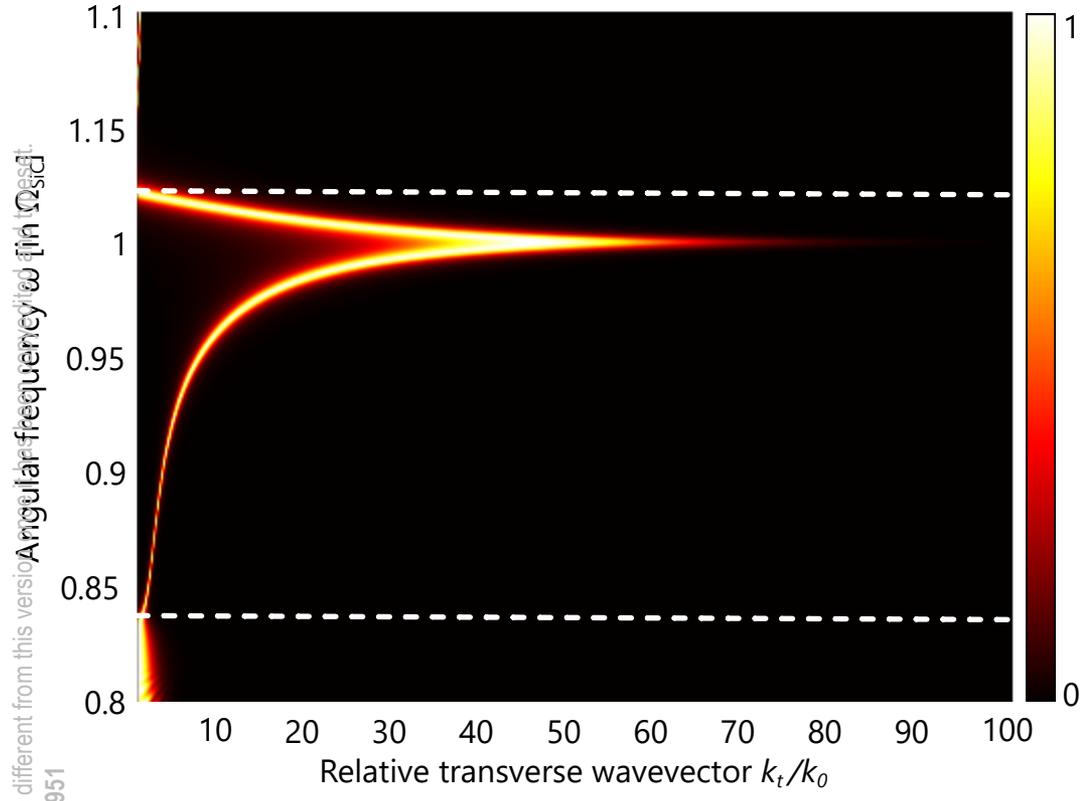

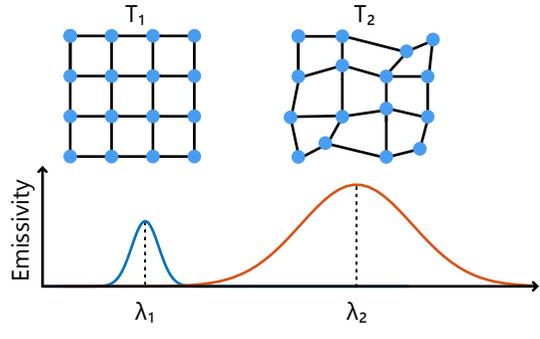

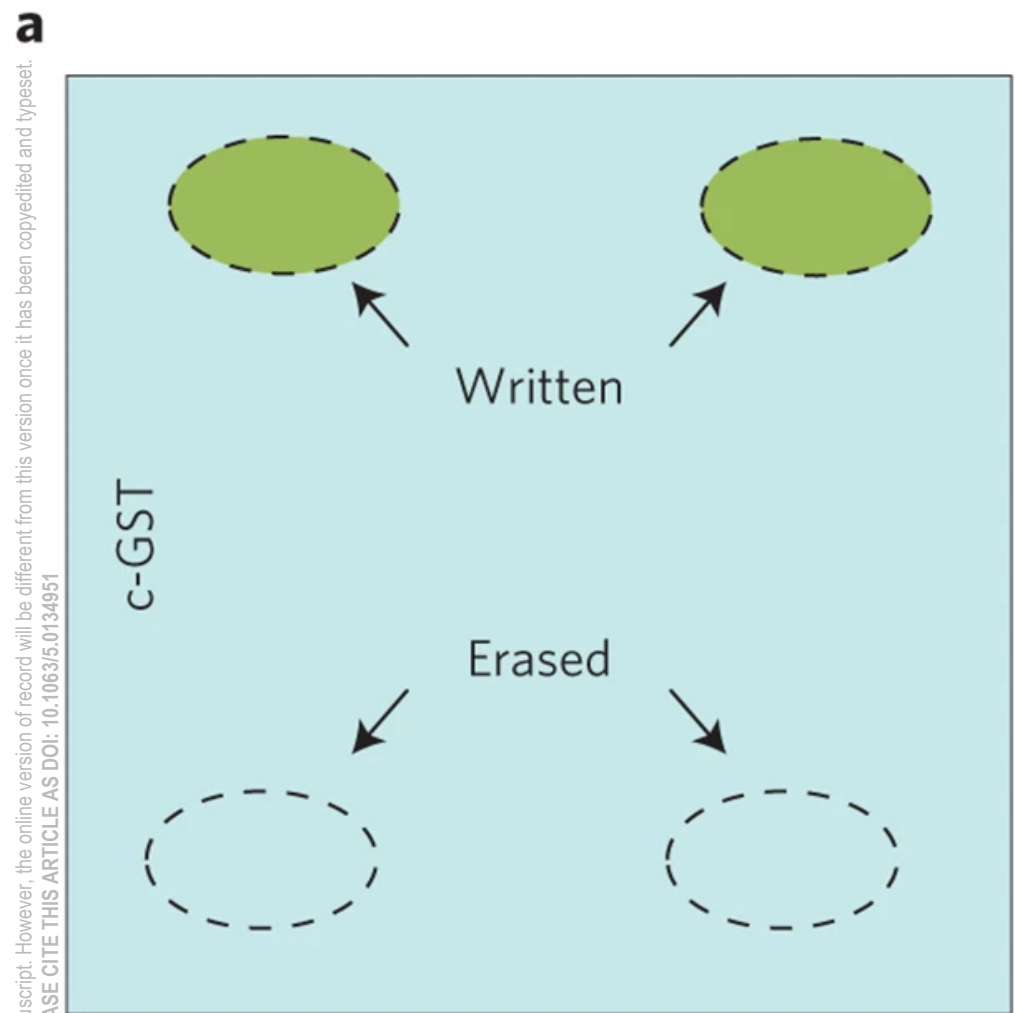

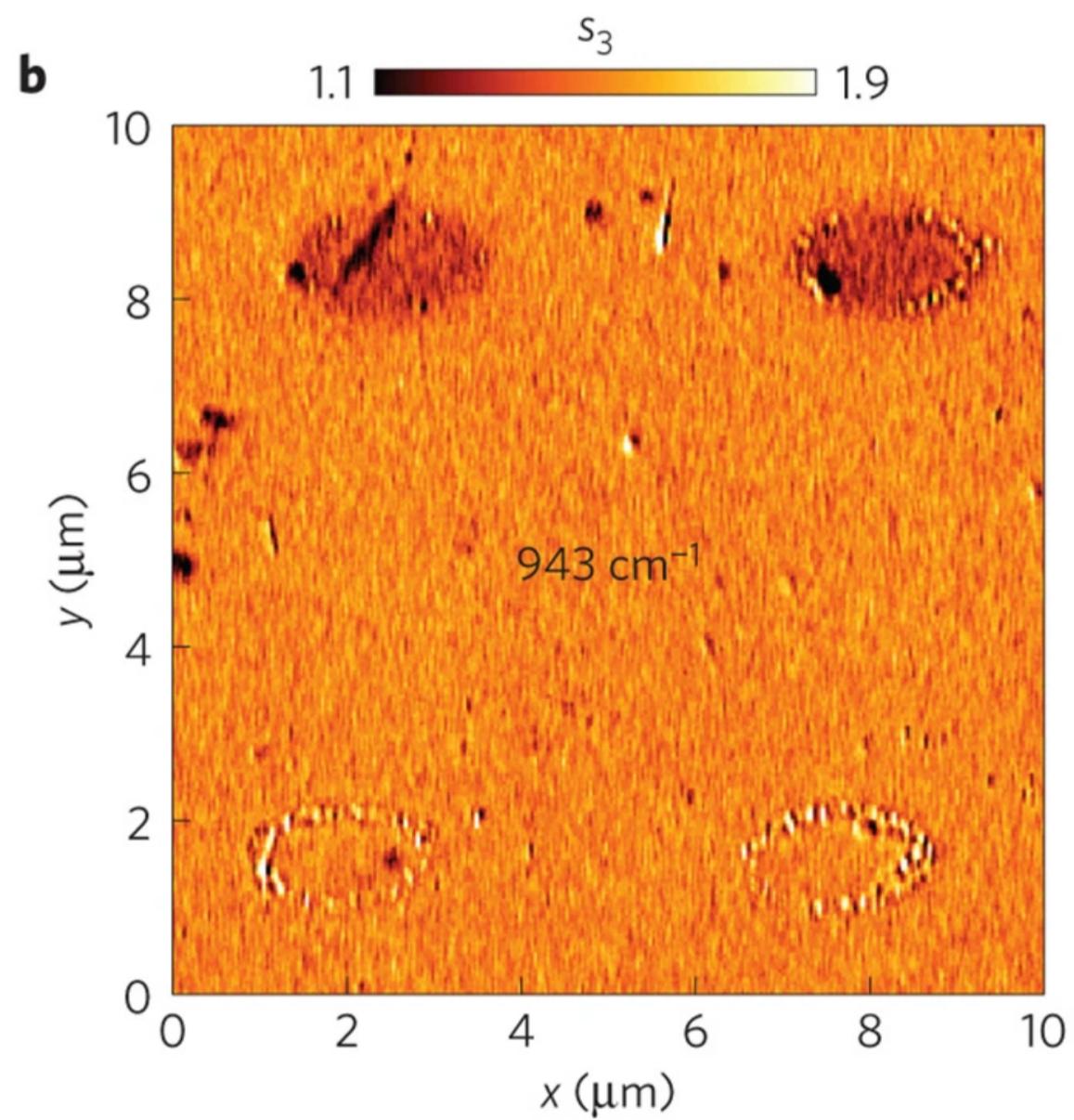

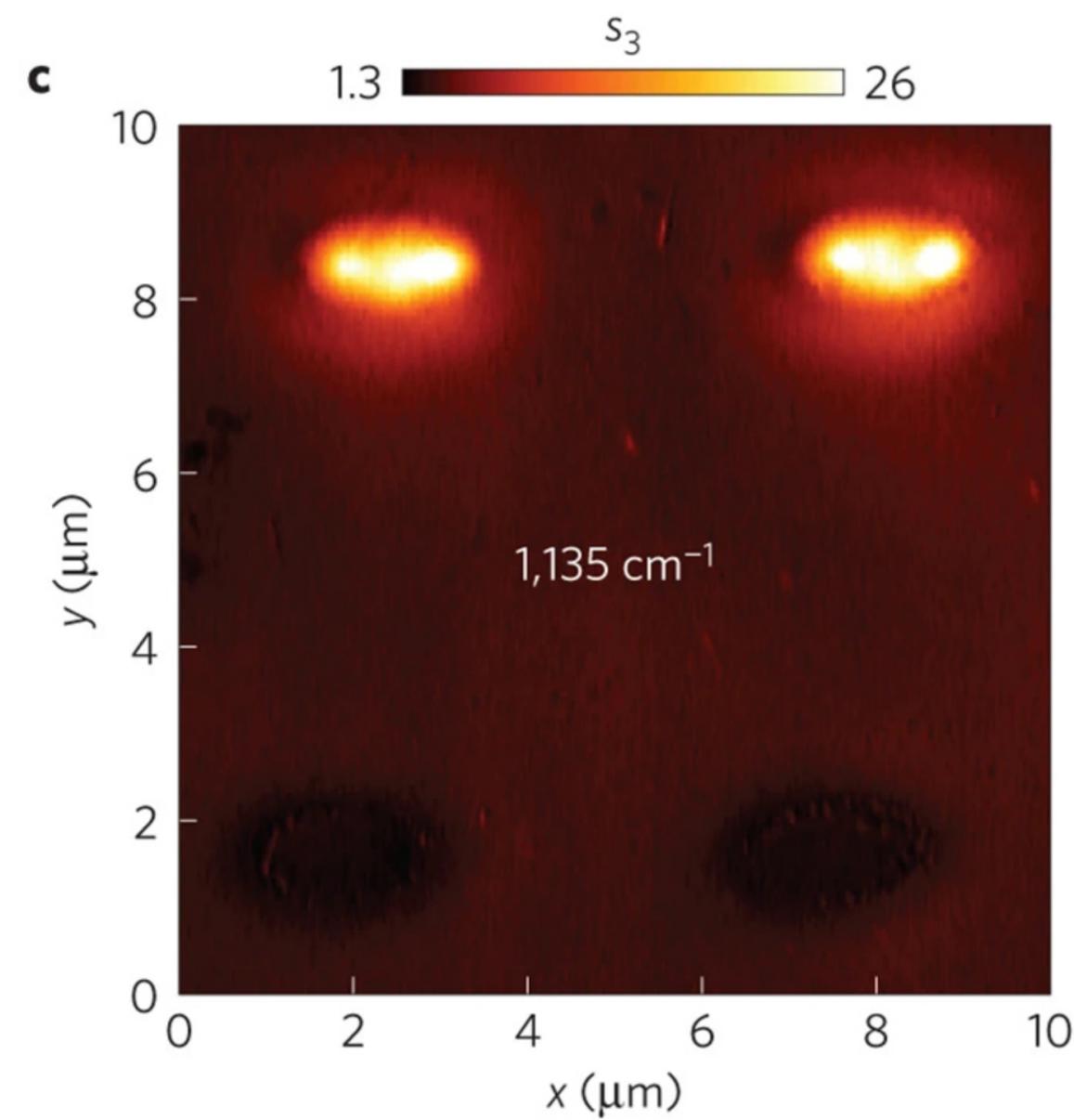

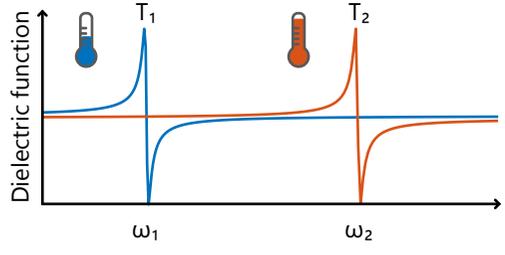

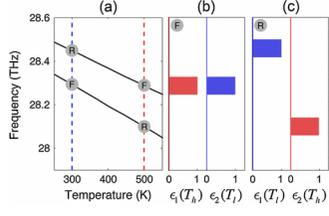

(a)

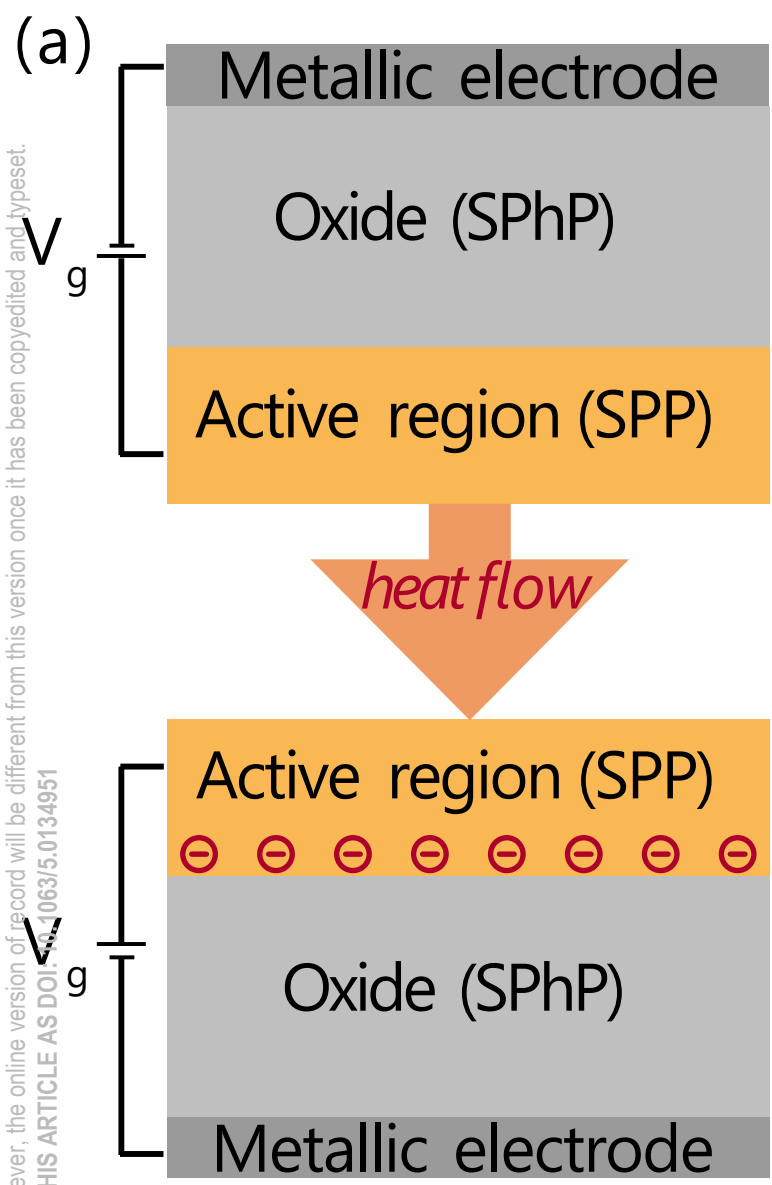

(b)

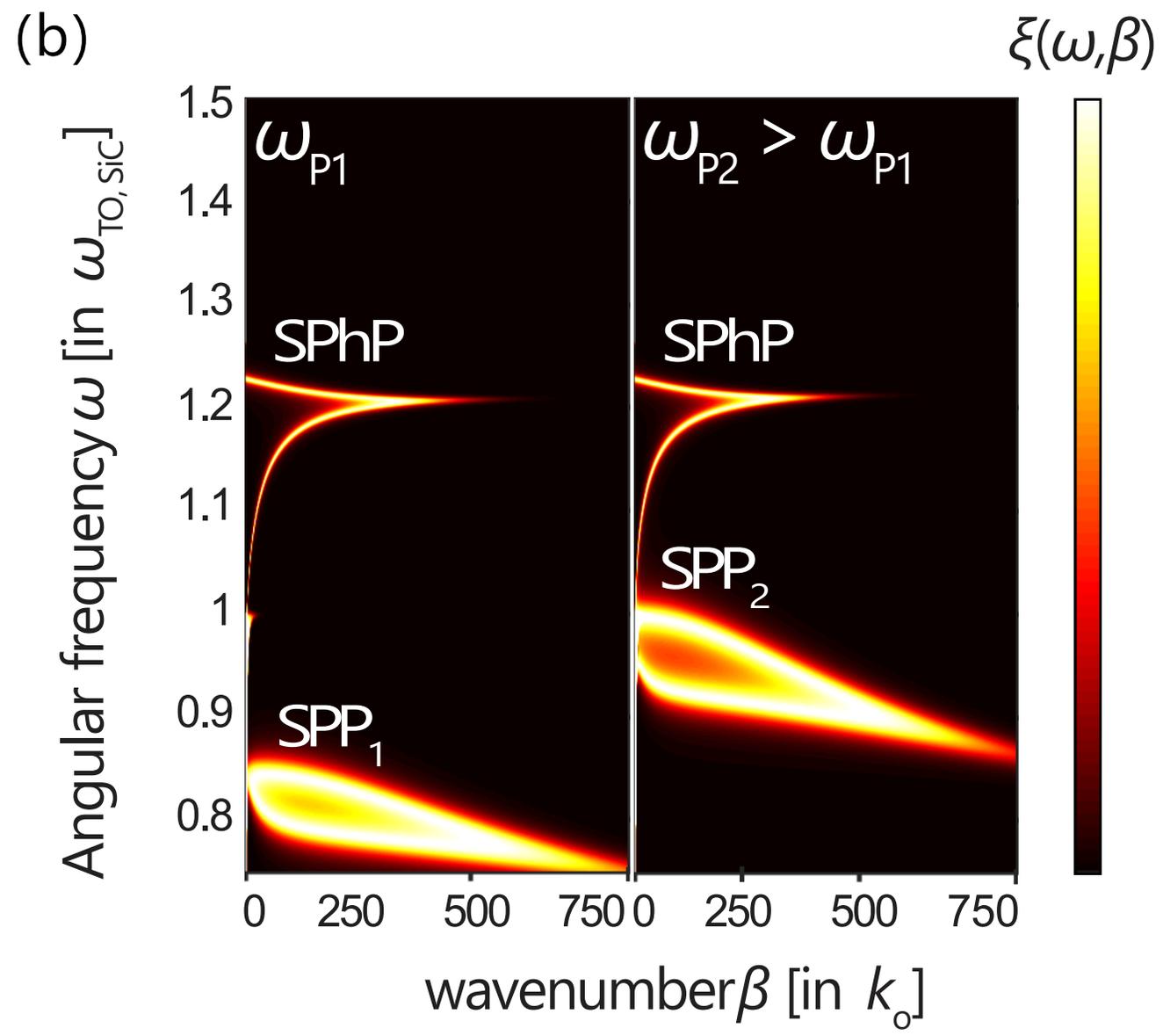

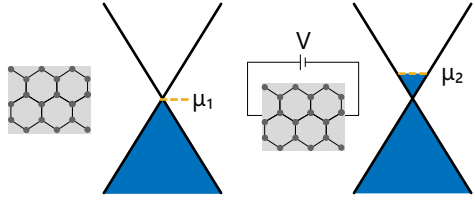

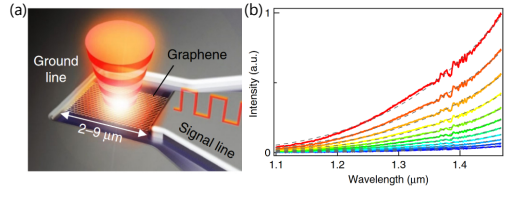

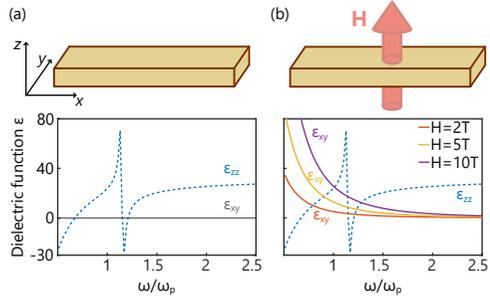

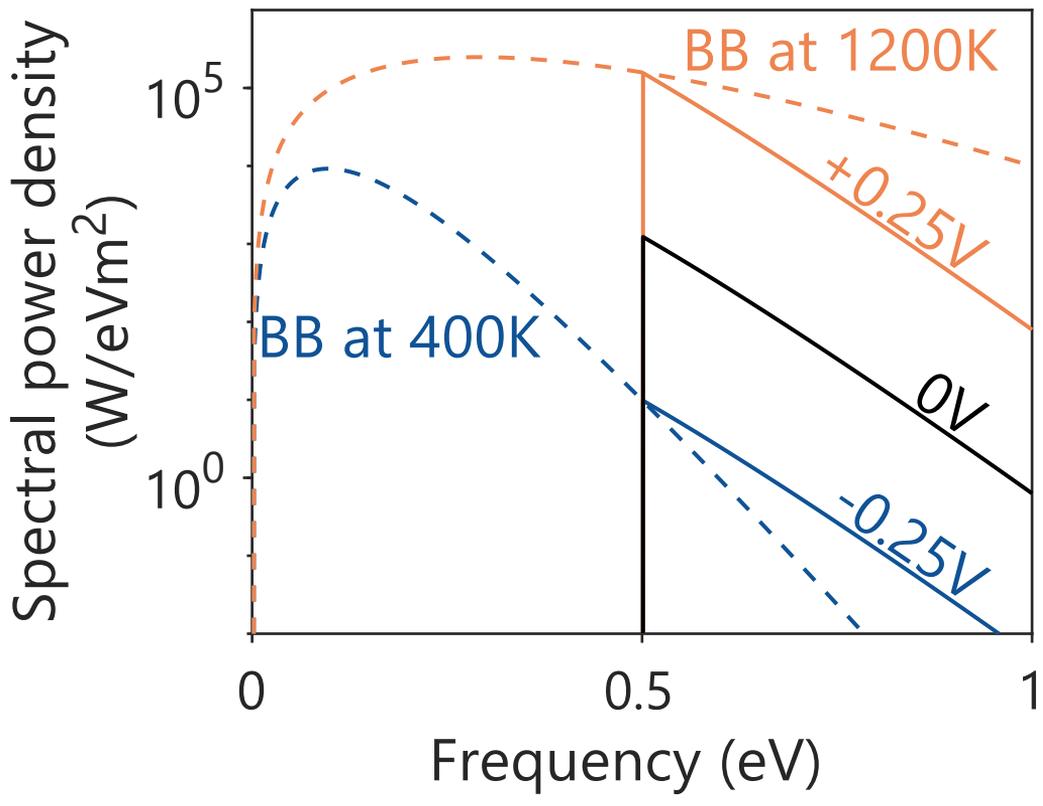